\input stromlo.tex

\title Dynamical evidence for black holes and dark halos

\author P.\ Tim de Zeeuw 

\shortauthor Tim de Zeeuw 

\affil Sterrewacht Leiden 

\abstract 
A variety of observational approaches have provided evidence for
extended halos of dark matter surrounding elliptical galaxies, and for
massive dark objects in (some of) their nuclei.  What are the
properties of the dark halos? How do they relate to the parameters of
the luminous matter? What is the demographics of massive central black
holes, and what is their dynamical role in shaping the host galaxy?
Recent work in this area is described, with special attention to the
construction of state-of-the-art dynamical models that incorporate all
available observational constraints. 
\index{dark matter}
\index{black holes}
\index{dynamical models}
\index{triaxiality}
\index{velocity profiles}

\section Introduction

One of the fundamental questions of astronomy is: {\it How did
galaxies form, and how do they evolve?} Elliptical galaxies are of
particular interest in this area, since their stellar populations are
among the oldest in the Universe. They display a rich variety of
physical properties. The inner tens of parsecs often contain stellar
and gaseous disks, possibly associated with massive central black
holes, as well as unresolved nuclear spikes, kinematically decoupled
structures, double nuclei, and puzzling asymmetries (e.g., Jaffe et
al.\ 1994; Lauer et al.\ 1995, 1996; Carollo et al.\ 1997).  The
nuclear properties appear to correlate with the global structure of
the parent galaxy.  For example, the luminosity-density profiles of
ellipticals approach a power-law form $\rho (r)\propto r^{-\gamma}$ at
small radii $r$, where the giant ellipticals (with M$_B < -22.0$) have
{\it shallow} cusps (mean $\gamma\sim 0.8$), while low-luminosity
ellipticals (with M$_B > -20.5$) have {\it steep} cusps (mean $\gamma
\sim 1.9$; Gebhardt et al.\ 1996). At intermediate luminosities both
types of nuclear profile occur. It is unclear whether these properties
were imposed by the galaxy formation process, or are the result of
subsequent evolution (possibly influenced by the environment), or
both: Does the formation of the nucleus determine the structure of the
galaxy, or is it a repository of low angular momentum material
collected over the life-time of the galaxy?  What is the role of
mergers, and of the physical processes they might trigger? Do they
destroy pre-existing steep cusps to make larger systems with shallower
cusps? Crucial issues are:

\item{(i)} What is the role of dark halos? Cosmological 
simulations suggest that dark matter halos have a mildly cusped
universal density profile, and that they have a range of triaxial
shapes. The luminous galaxy forms by infall of baryonic material. This
results in a contraction of the dark halo, and a modification of its
shape.  What is the relation between the profile and the shape of the
contracted dark halos and the properties of the luminous galaxy in it?

\item{(ii)} What is the role of a central massive dark object? 
Do all ellipticals contain one? Is this a black hole? What is the
distribution of black hole masses? How does the black hole mass relate
to the global/nuclear properties of the parent galaxy?  What is the
dynamical relevance of the black hole? Does it drive the galaxy
towards axisymmetry from the inside out?

\noindent
Observational and theoretical efforts are needed to answer these
fundamental questions.  Two-dimensional spectroscopy over the entire
optical extent of the galaxies is required, complemented by high
spatial resolution imaging and spectroscopy of the nuclei. In
addition, the construction of flattened multi-component dynamical
models that have the full variety of anisotropic velocity
distributions is essential for a correct interpretation of this data.

\section Determination of unseen mass

Studies of the dark matter content of elliptical galaxies concentrate
on the properties of (i) dark halos surrounding the luminous galaxies,
and (ii) massive dark objects such as black holes in their nuclei. In
both cases we need to measure the {\it total} mass-to-light ratio
$M/L$ as a function of position, and then to compare it with the {\it
stellar} $M/L$ of the luminous matter. This requires several steps.

\subsection Intrinsic shapes

The luminosity distribution $L$ is usually found by deprojecting the
surface brightness distribution. This deprojection is non-unique, and
requires the determination of the intrinsic shape for the galaxy, and
of the direction of viewing. The mass distribution $M$ is derived from
the total gravitational potential. In some cases this can be inferred
from the emission of extended X-ray gas (e.g., Forman, Jones \& Tucker
1985; Buote \& Canizares 1996), from gravitational lensing (Maoz \&
Rix 1993; Kochanek 1995), or from the kinematics of cold gas at large
(Franx et al.\ 1994) or small (Harms et al.\ 1994) radii. In general,
however, the potential must be derived by dynamical modeling of the
stellar kinematics, either obtained from integrated light
measurements, or from radial velocities of individual objects such as
planetary nebulae and globular clusters. This again requires the
determination of the intrinsic shape and of the viewing direction.

\subsection Orbital structure

Giant elliptical galaxies very likely have stationary, or at most
slowly tumbling, triaxial figures.  The orbital structure of (even
mildly) triaxial systems is rich, and depends on the tumbling rate,
the degree of triaxiality, and the central mass concentration.
Theoretical work carried out in the past two decades has established
that dynamical models with shallow cusps can be constructed for a wide
range of triaxial shapes (see the review by de Zeeuw 1996, and
references therein). The stars in such models may occupy box orbits,
various families of tube orbits, minor orbit families, or irregular
orbits. There are many ways to populate the different orbits so as to
reproduce the same triaxial shape. The resulting dynamical models
therefore differ in their velocity distributions. The regular box
orbits disappear in models with steep cusps, and the minor orbit
families as well as the irregular orbits become more important
(Gerhard \& Binney 1985; Miralda-Escud\'e \& Schwarzschild 1989).
This change in orbital structure may limit the degree of triaxiality
of equilibrium models for low-luminosity ellipticals: few or no orbit
combinations can reproduce a steep-cusped strongly triaxial model
(Kuijken 1993; Schwarzschild 1993; Merritt \& Fridman
1996). Similarly, it is suspected that strongly triaxial models with
massive central black holes can not be in dynamical equilibrium.
However, much is still to be learned about the allowed degree of
non-axisymmetry.

\subsection Measuring the velocity anisotropy

The anisotropy of the velocity field must be measured in order to
distinguish its variations from true $M/L$ variations caused by unseen
matter (e.g., Binney \& Mamon 1982; Gerhard 1993). Recently,
techniques have been developed to measure not only the mean streaming
velocities $\langle v \rangle$ and velocity dispersions $\sigma$, but
also the shape of the entire line-of-sight velocity distribution
(velocity profile VP; e.g., Rix \& White 1992; van der Marel \& Franx
1993). The VP shapes constrain the anisotropy of the velocity
distribution. These measurements require high signal-to-noise high
spectral resolution spectroscopic data, which can now be obtained
routinely (e.g., van der Marel et al.\ 1994; Carollo et al.\ 1995;
Statler et al.\ 1996).

\subsection Dynamical modeling 

Finally, fully general dynamical models are required to derive the
mass distribution from the measured VPs. These models must: (i) not
require the availability of analytic integrals of motion; (ii) put no
restrictions on the form of the gravitational potential, and thus
allow for arbitrary geometry and the inclusion of a central black
hole; (iii) allow for multiple luminous components such as nuclear
disks and kinematically decoupled structures; (iv) put no restriction
on the form of the velocity distribution, and thus allow exploration
of the full range of velocity anisotropy. Advances in computer
technology make it now possible to construct such models, as discussed
below.

\section An extension of Schwarzschild's Method

A versatile method for building galaxies was introduced by
Schwarzschild (1979, 1982), who used it to establish the existence of
triaxial equilibrium models for galaxies with finite density cores,
stationary as well as tumbling.  In the original version of the
method, stellar orbits were integrated numerically in a chosen galaxy
potential, and then populated so as to reproduce the associated
density distribution. Richstone (1980, 1984) built scale--free
axisymmetric models with this technique. In the past decade the method
has been used to build a variety of spherical, axisymmetric and
triaxial galaxy models which also include the observed velocity
dispersions as constraints (e.g., Richstone \& Tremaine 1984, 1985;
Pfenniger 1984; Levison \& Richstone 1985; Zhao 1996).

\figureps[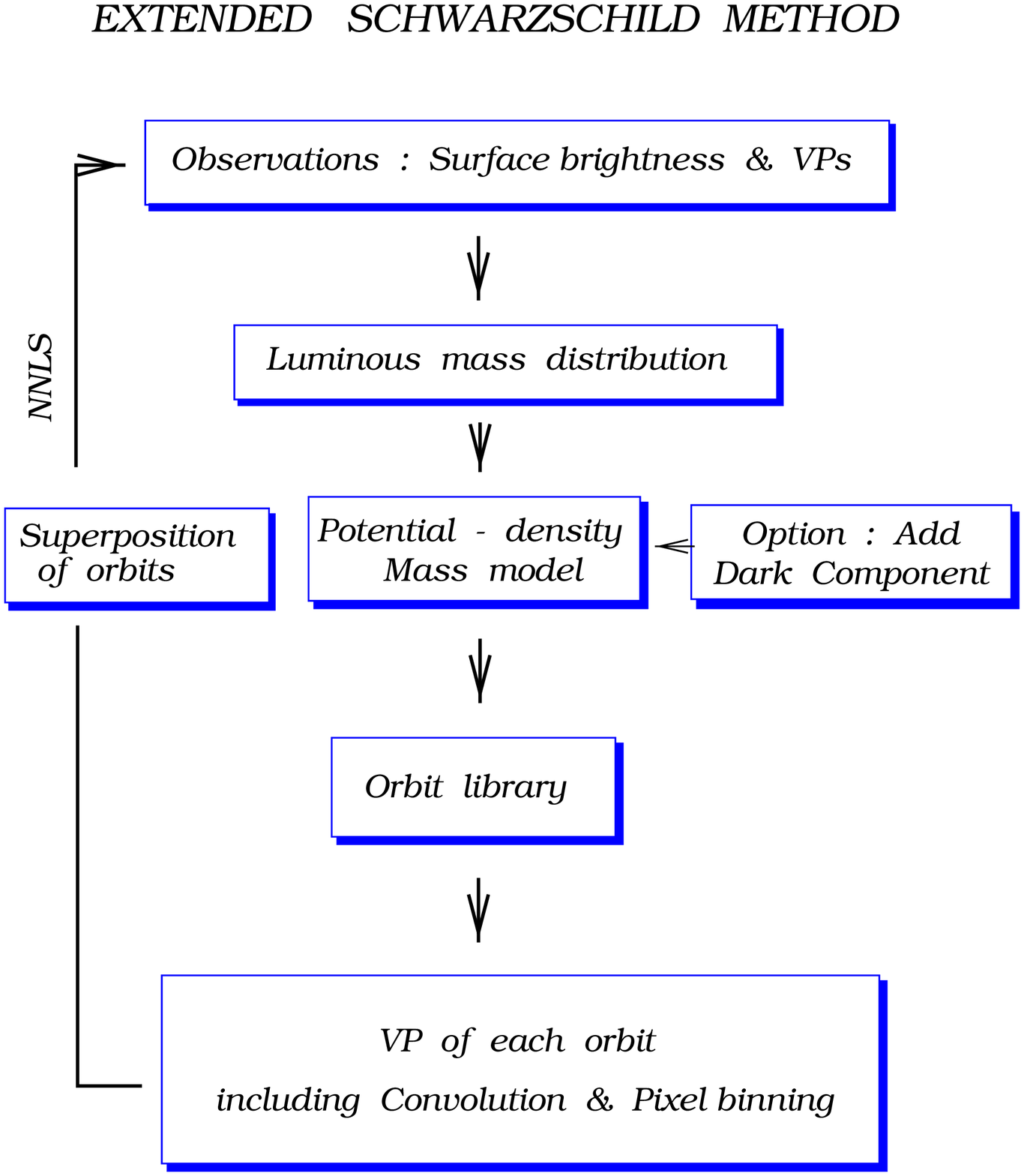,0.8\hsize] 1.\ Flowchart of the extended version 
of Schwarzschild's numerical orbit superposition method for the
construction of fully general dynamical models for galaxies. The
velocity profiles (VPs) of each orbit are fitted to the observed
surface brightness and kinematic measurements by means of a
non-negative least squares (NNLS) algorithm.

We have recently extended the Schwarzschild method further by
incorporating the entire VPs as constraints in the construction of
fully anisotropic models with arbitrary geometry. We are therefore
finally in the position to quantify the velocity distribution in
ellipticals, and to measure their dark matter content. We have
completed the spherical version of the code (Rix et al.\ 1997), as
well as the axisymmetric version (Cretton et al.\ 1997; van der Marel
et al.\ 1997c). A similar code for tumbling geometry, and for lopsided
systems, is under construction (Zhao et al.\ in preparation). A
further extension will include velocity measurements of individual
objects (e.g., planetary nebulae or globular clusters).

Our version of Schwarzschild's method consists of the following steps,
illustrated in Figure 1: (i) a mass density is chosen that fits the
galaxy photometry after projection; (ii) the gravitational potential
is calculated, including the contribution of dark components such as a
nuclear point mass or an extended halo; (iii) a representative sample
of orbits is calculated; (iv) the contribution of each orbit to the
observed surface brightness and the observed VPs is calculated.  These
orbital VP contributions are convolved with the appropriate Point
Spread Function, and aperture binned; (v) a non-negative least-squares
(NNLS) fit is performed to determine the combination of orbits that
reproduces the photometric and kinematic observations, taking into
account the observational errors. A smoothness criterion is used to
exclude solutions with very uneven populations of neighboring orbits.

Examples of orbital VPs can be found in Cretton's contribution to this
volume.  Our implementation uses the Gauss-Hermite expansion of the
VPs of the orbits and of the model, i.e., we calculate the orbital
contributions to the mean rotation velocity $\langle v \rangle$, to
the velocity dispersion $\sigma$, and to the parameters $h_i$ (with
$i=3, 4, \ldots$) which quantify the non-Gaussian shape of the VP (van
der Marel \& Franx 1993).  Other ways of parametrizing the VPs can be
used as well, including straight binning in the velocity coordinate,
as long as this yields linear constraints. The Gauss-Hermite expansion
uses a modest number of parameters, and hence does not require
excessive cpu time and storage space.

The applications of this machinery are many. Below we briefly
illustrate its power in constraining the mass distribution in
ellipticals by considering the case of the massive black hole in the
nucleus of M32, and the case of the dark halo around the E0 galaxy NGC
2434. More extensive descriptions of these applications can be found
in the contributions to this volume by van der Marel and by Rix,
respectively.

\section The black hole in M32 

The search for central massive dark objects in the nuclei of quiescent
elliptical galaxies has received much attention in the past two
decades, with a steady improvement in the quality of the data, and in
the sophistication of the modeling techniques (Kormendy \& Richstone
1995).  In order to find stellar dynamical evidence for a massive
central dark object, we need to probe inside the radius where the
stellar motions are dominated by the gravitational field of the dark
object. This radius of influence is typically of the order of $1''$ or
less, so spectroscopic observations at HST angular resolution (e.g.,
Kormendy et al.\ 1996; Richstone, van den Bosch, van der Marel, all
elsewhere in this volume), or better, are required.

M32 is a low-luminosity E3 galaxy \index{M32} with a steep stellar
cusp. The inner region shows no minor axis rotation and no isophote
twist.  This is circumstantial evidence that M32 is likely to be
nearly axisymmetric. It has been suspected of harboring a black hole
for over a decade, but so far all mass determinations were based on a
comparison of ground-based data with either spherical models with
general velocity distributions, or with flattened models with special
distribution functions (Tonry 1987; Dressler \& Richstone 1988; van
der Marel et al.\ 1994; Qian et al.\ 1995; Bender, Kormendy \& Dehnen
1996).

We have recently obtained {\tt FOS} spectra at 8 positions along the
major axis of M32, within the inner $0.5''$. We have applied the
extended Schwarzschild method to build axisymmetric models that match
the combined ground-based and {\tt FOS} kinematics (van der Marel et
al.\ 1997a, b, c). Our modeling follows the scheme of Figure~1, where
the dark potential is taken to be either the Keplerian potential of a
point mass (to describe a massive black hole), or that of a model with
a finite scale-length, such as a Plummer potential or a cusped model
(to describe a dense cluster of dark objects, see e.g., Gerhard 1994).

\figuretwops[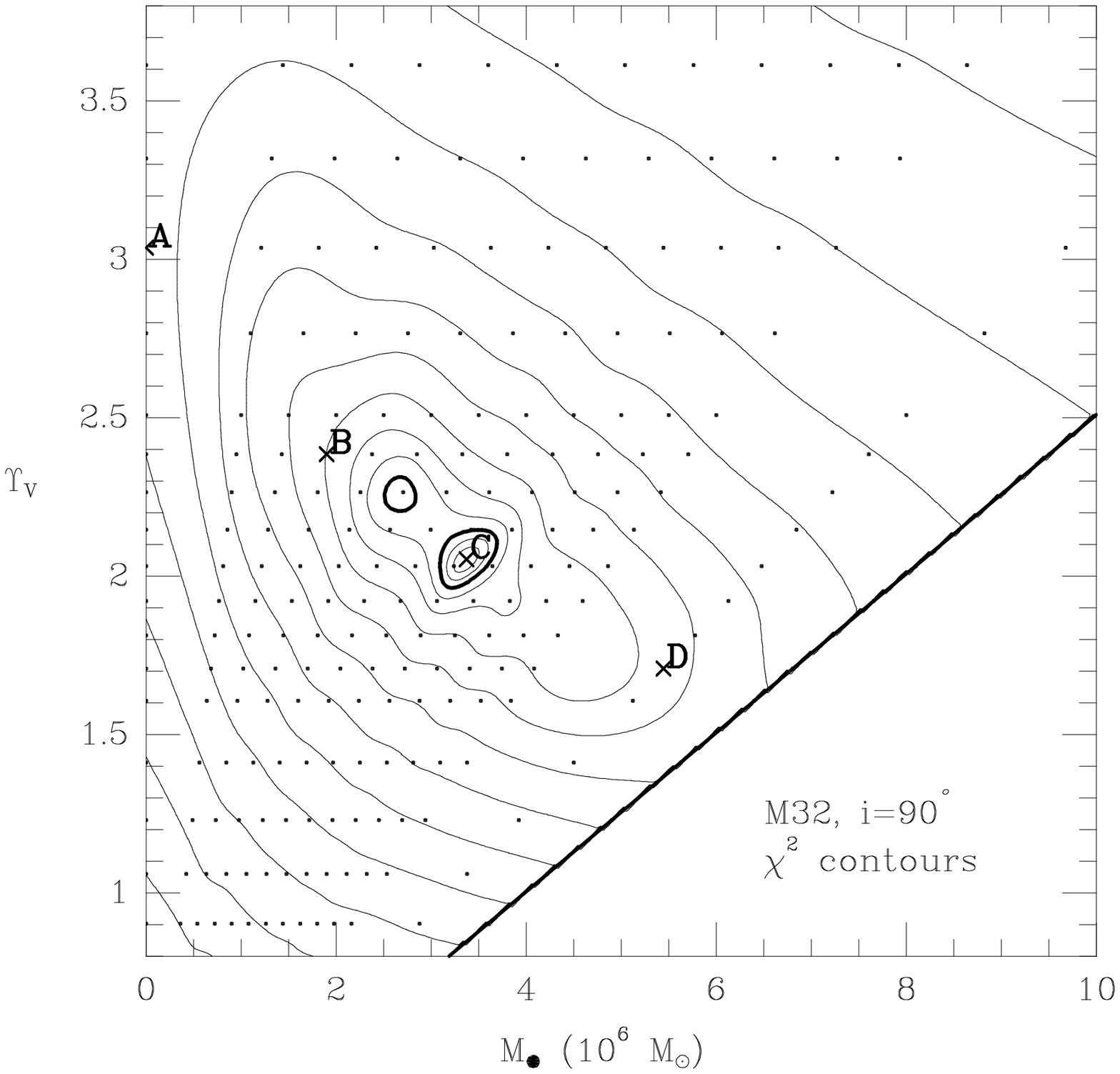,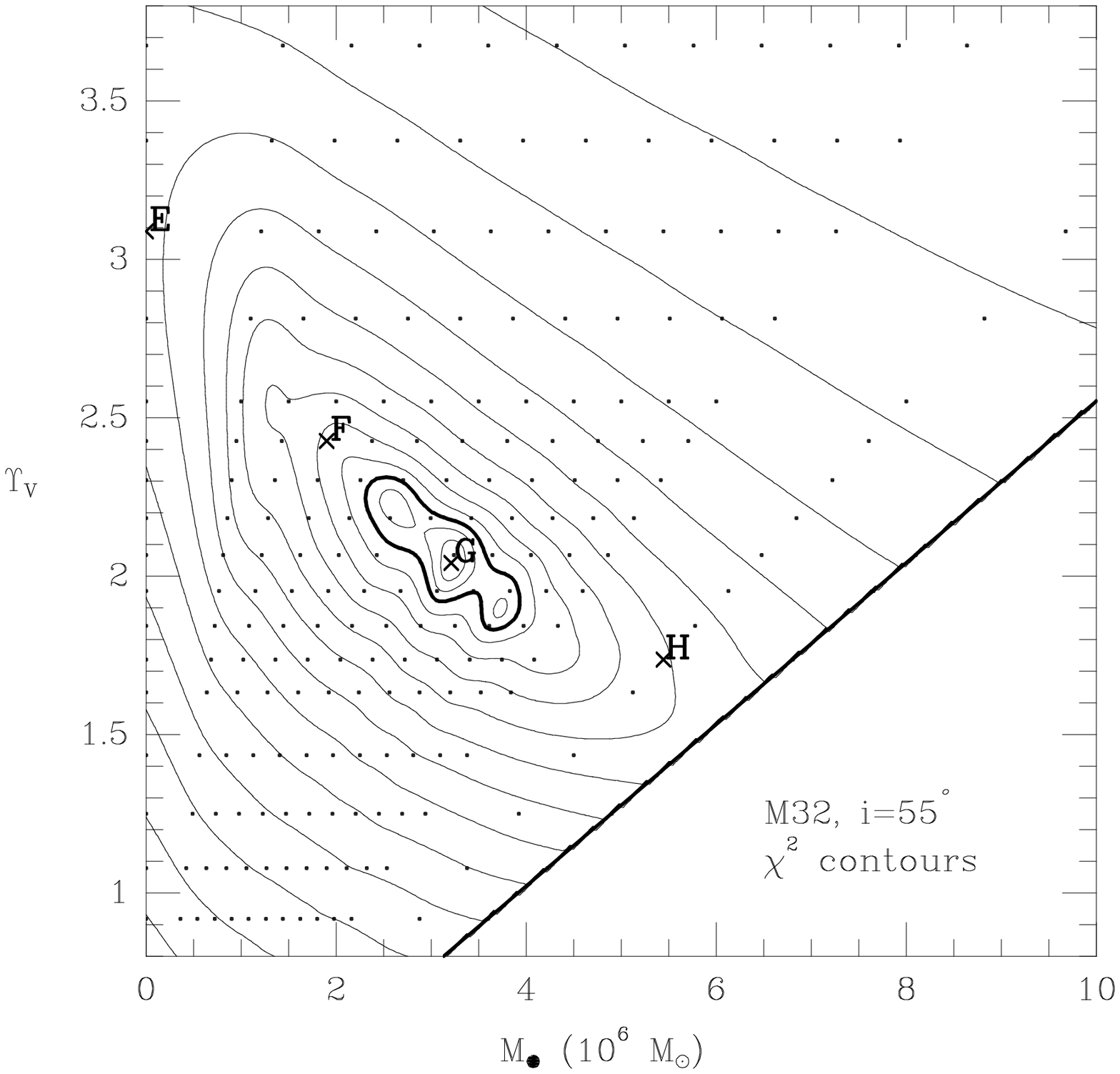,0.45\hsize] 2. Contours of 
the $\chi^2$ that measures the quality of fit to groundbased and {\tt
FOS} kinematic data for M32, for models with a fully general dynamical
structure, in the plane of $M_{\rm BH}$ and stellar mass-to-light
ratio $\Upsilon_V$ (in solar units).  The dots indicate dynamical
models that were calculated.  The dynamical structure and predicted
kinematics of the models labeled A-H are discussed in van der Marel et
al (1997c).  The first three (heavy) contours define the formal
68.3\%, 95.4\% and 99.73\% confidence regions. a) edge-on model; b)
inclined model. The case of no central dark mass is firmly ruled out:
the mass of the central object lies in the range $(3.4\pm1.6) \times
10^6 M_\odot$.

The models have three free parameters: the V-band mass-to-light ratio
$\Upsilon_V$ of the stellar population, the dark mass $M_{\rm BH}$,
and the inclination $i$. However, we need to calculate only one orbit
library for models with the same value of $M_{\rm BH}/\Upsilon_V$,
since the potentials of such models are identical except for a
normalization of the velocity scale. Each $(M_{\rm BH},
\Upsilon_V)$ combination does instead require a separate NNLS fit. 

All models at a chosen inclination are constructed for the same
parametrized luminous density distribution, which is fixed by fitting
its projection to the surface brightness distribution of M32,
determined from ground-based and HST data (Lauer et al.\ 1992).  If
M32 is axisymmetric, the deprojection of the surface brightness
distribution is unique for $i=90^\circ$, but is non-unique for other
inclinations (Rybicki 1987). So, our parametrization of the luminous
density may not be the right model if $i\not=90^\circ$, even though it
fits the projection. However, van den Bosch (1997) has used the
formalism of konus densities introduced by Gerhard \& Binney (1996)
and Kochanek \& Rybicki (1996), and concluded that non-uniqueness in
the deprojection of M32 is negligible.

Because the measurement errors are taken into account in the NNLS fit
to the data, we are able to compare the different dynamical models in
an objective way, by computing the $\chi^2$ of each fit. The result is
illustrated in Figure 2.  The left panel shows the $(M_{\rm BH},
\Upsilon_V)$ plane for the edge-on case $i=90^\circ$.  The right hand
panel shows the same diagram for $i=55^\circ$.  In both panels there
is a clear minimum in the $\chi^2$ contours, for $M_{\rm BH} \sim 3.4
\times 10^6 M_\odot$.  Representative orbits in the best fitting
edge-on model are illustrated in Cretton's contribution to this
volume. Based on extensive simulations of observational errors, and of
discretization errors in the numerical procedure, the conservative
estimate is that the allowed range for $M_{\rm BH}$ is $(3.4\pm 1.6)
\times 10^6 M_\odot$, independent of the inclination. The modeling
also puts an upper limit on the dimensions of the dark mass, and rules
out that the central object is an extended cluster of dark
remnants. This finally puts the presence of a black hole in M32 beyond
doubt!

\section The dark halo of NGC 2434

Even though the existence of dark matter around elliptical galaxies is
established, it remains unclear whether the stellar and dark mass are
as tightly coupled as they appear to be in spiral galaxies, where they
conspire to give a flat rotation curve, and how the dark halo
properties correlate with those of the luminous galaxy. Part of the
difficulty comes from our ignorance about the form of the dark
potential.

We have used our version of Schwarzschild's method to measure the dark
matter content of the E0 galaxy NGC 2434. \index{NGC 2434} We have
firmly ruled out constant $M/L$ models for this galaxy, regardless of
the orbital anisotropy. Furthermore, we have considered the
cosmologically motivated `star+halo' potentials of Navarro et al.\
(1996; see also Cole \& Lacey 1996), modified so as to incorporate the
accumulation of baryonic matter under the assumption of adiabatic
invariance. These `star+halo' potentials provide an excellent fit to
the data. The best fitting potential has a circular velocity that is
constant to within $\sim$ 10\% between 0.2 and 3 effective
radii. Roughly half of the mass inside one effective radius is found
to be dark.

We are currently extending this program to study the dark matter
properties of a statistically significant number of ellipticals,
covering a large range in luminosity.

\section The next steps 

Major progress in this area is just around the corner. On the
observational side, the kinematics of the bright nuclear regions are
being studied at the appropriate resolution to measure the signature
of massive central black holes, especially in quiescent galaxies. {\tt
STIS} on HST will improve the efficiency of these studies by providing
long-slit spectra with 0.1$''$ resolution.  In addition, {\it
two-dimensional} (integral field) spectroscopy is finally unveiling
the rich structure of the velocity fields and line-strength
distributions of triaxial and multi-component systems.  For the
investigation of the nuclei, integral field spectrographs with small
fields of view but high spatial resolution (0.1--0.25$''$, by use of
adaptive optics) are currently being constructed at a number of major
observatories, including {\tt OASIS} at the Canada-France-Hawaii
Telescope, and {\tt GMOS} for Gemini. For studies of the dark halos,
intermediate spatial resolution spectroscopy over the entire optical
extent of elliptical galaxies is needed. To this end, the Observatoire
de Lyon, the Astronomy group at the University of Durham, and the Leiden
Observatory have recently formed the {\tt SAURON} Consortium, with
PI's Bacon, Davies \& de Zeeuw, and co-I's Carollo, Emsellem, Monnet
\& Allington-Smith, to build the integral-field spectrograph {\tt
SAURON} (Spectroscopic Areal Unit for Research on Optical Nebulae),
for use on a 4m class telescope such as the WHT on La Palma. {\tt
SAURON} uses an array of hexagonal lenslets that provides nearly 1500
simultaneous spectra at $1''$ resolution in a $30'' \times 35''$ field
of view with a separate option for accurate sky-subtraction (120
spectra at $2.2'$ from the field center). The instrument also has a
high spatial resolution mode (see Table 1). {\tt SAURON} will allow us
to measure the full, complex, kinematics and metallicity distributions
of the giant elliptical galaxies.  The extended Schwarzschild method
will be used to model these data.  The dynamical models will be
constructed from orbits that not only cover phase-space, but also come
in different flavors, i.e., different values of the line-strength. The
extra constraints provided by the two-dimensional line-strength
observations will then allow the determination of the intrinsic
physical properties of the stellar populations as a function of
position in the galaxy.

\table 1. Specifications of SAURON

@l @c @c \\
\tworules 
&Spatial sampling      &0.3$''$              &1.0$''$            \\
&Field of view         &$10.3'' \times 9.0''$  &$34.5'' \times 29.9''$ \\
&Spectral resolution   &65 km/s               &73 km/s             \\
&Spectral sampling     &1.3 \AA/pixel        &1.3 \AA/pixel      \\
&Spectral range        &4820--5410 \AA       &4820--5410 \AA     \\
\onerule
\space\space

On the theoretical side, it is essential to investigate further the
question of the existence of triaxial equilibria as a function of
shape, cusp slope, tumbling rate, and presence of a central dark mass
or a dark halo. Furthermore, the central stellar density in some
ellipticals with a cusp is so high that the two-body relaxation time
for stellar encounters becomes significantly less than a Hubble time,
so that the collisionless approximation fails sufficiently close to
the center. This has as yet received little attention in the context
of galactic nuclei. An accurate treatment of secular evolution, such
as driven by the growth of a central black hole, is the next frontier
for N-body simulations. Issues to be addressed include the role of the
central black hole in determining the cusp slope and shape of the host
galaxy, and the effect of the capture of another galaxy, possibly with
its own black hole.

\acknowl 
It is a pleasure to thank Marcella Carollo, Nicolas Cretton,
Hans--Walter Rix, Roeland van der Marel, and Hongsheng Zhao for
enjoyable collaborations, for permission to quote from joint work, and
for comments on the manuscript.

\references

Bender R., Kormendy J., Dehnen W., 1996, ApJ, 464, L123

Binney J.J., Mamon G.A., 1982, MNRAS, 200, 361

Buote D.A., Canizares C.R., 1996, ApJ, 468, 184

Carollo C.M., de Zeeuw P.T., van der Marel R.P., Danziger I.J., Qian E.E., 
1995, ApJL, 441, 25

Carollo M., Franx M., Illingworth G.D., Forbes D., 1997, ApJ, 481, in press

Cole S., Lacey C., 1996, MNRAS, 281, 716

Cretton N., de Zeeuw P.T., van der Marel R.P., Rix H.W., 1997, ApJ, submitted 

de Zeeuw P.T., 1996, in Gravitational Dynamics, eds O.\ Lahav, E.\ \& R.J.\
Terlevich (Cambridge Univ.\ Press), 1

Dressler A., Richstone D.O., 1988, ApJ, 324, 701

Forman C., Jones C., Tucker W., 1985, ApJ, 293, 102

Franx M., van Gorkom J., de Zeeuw P.T., 1994, ApJ, 436, 642

Gebhardt K., et al., 1996, AJ, 112, 105

Gerhard O.E., 1993, MNRAS, 265, 213

Gerhard O.E., 1994, in The Nuclei of Normal Galaxies, eds R.\ Genzel
\& A.I.\ Harris (Dordrecht: Kluwer), p.\ 267

Gerhard O.E., Binney J.J., 1985, MNRAS, 216, 467

Gerhard O.E., Binney J.J., 1996, MNRAS, 279, 993

Harms R.J., et al., 1994, ApJ, 435, L35

Jaffe W., Ford H.C., O'Connell R.W., van den Bosch F.C., Ferrarese L.,
1994, AJ, 108, 1567

Kochanek C., 1995, ApJ, 445, 559

Kochanek, C., Rybicki G.B., 1996, MNRAS, 280, 1257

Kormendy J., Richstone D.O., 1995, ARAA, 33, 581 

Kormendy J., et al., 1996, ApJ, 459, L57

Kuijken K., 1993, ApJ, 409, 68 

Lauer T.R., et al., 1992, AJ. 104, 552

Lauer T.R., et al., 1995, AJ, 110, 2622

Lauer T.R., et al., 1996, ApJ, 471, L79

Levison H.F., Richstone D.O., 1985, ApJ, 295, 349

Maoz D., Rix H.-W., 1993, ApJ, 416, 425

Merritt D.R., Fridman T., 1996, ApJ, 460, 136

Miralda-Escud\'e J., Schwarzschild M., 1989, ApJ, 339, 752

Navarro J., Frenk C., White S.D.M., 1996, ApJ, 462, 563

Pfenniger D., 1984, A\&A, 141, 171

Qian E., de Zeeuw P.T., van der Marel R.P., Hunter C., 1995, MNRAS,
274, 602

Richstone D.O., 1980, ApJ, 238, 103

Richstone D.O., 1984, ApJ, 281, 100

Richstone D.O., Tremaine S.D., 1984, ApJ, 286, 27

Richstone D.O., Tremaine S.D., 1985, ApJ, 296, 370

Rix H.W., White S.D.M., 1992, MNRAS, 254, 389

Rix H.W., de Zeeuw P.T., Carollo C.M., Cretton N., van der Marel R.P., 1997, 
ApJ, submitted

Rybicki G.B., 1987, in IAU Symposium 127, Structure and Dynamics of
Elliptical Galaxies, ed.\ P.T.\ de Zeeuw (Dordrecht: Kluwer), p.\ 397

Schwarzschild M., 1979, ApJ, 232, 236

Schwarzschild M., 1982, ApJ, 263, 599 

Schwarzschild M., 1993, ApJ, 409, 563

Statler T., Smecker-Hane T., Cecil G., 1996, AJ, 111, 151

Tonry J.L., 1987, ApJ, 322, 632
 
van den Bosch F.C., 1997, MNRAS, in press 

van der Marel R.P., Franx M., 1993, ApJ, 407, 525

van der Marel R.P., Rix H.-W., Carter D., Franx M., White S.D.M., de
Zeeuw P.T., 1994, MNRAS, 268, 521

van der Marel R.P., de Zeeuw P.T., Rix H.W., Quinlan G.D., 1997a,
Nature, in press.

van der Marel R.P., de Zeeuw P.T., Rix H.W., 1997b, submitted to ApJ

van der Marel R.P., Cretton N., de Zeeuw P.T., Rix H.W., 1997c,
submitted to ApJ

Zhao H.S., 1996, MNRAS, 278, 488

\bye